\documentclass[11pt]{article}
\usepackage{jheppub}
\usepackage{pstool}
\pdfoutput=1
\usepackage{color}
\usepackage{float}
\usepackage{array}
\usepackage{amssymb}
\usepackage{amsmath}
\usepackage{amsthm}
\usepackage{graphicx}
\usepackage{caption}
\usepackage[labelsep=quad]{subcaption}
\usepackage{epstopdf}
\usepackage{placeins}
	
\usepackage{epsfig}
\usepackage{simpler-wick}
\def\pb[#1,#2]{\{#1, #2\}}
\def\deb[#1,#2]{[#1,#2]_{\text{D.B.}}}

\def\Or[#1]{{\text{O}}\left({#1}\right)}
\def\dotl[#1,#2]{\left\langle #1,\, #2 \right\rangle}
\def\dotlb[#1,#2]{\left\langle #1,\, #2 \right\rangle}
\def\dotlm[#1,#2]{\left[ #1,\, #2 \right]}
\def\dotp[#1,#2]{(\vect{#1} \cdot\vect{#2})}
\def\aff[#1,#2]{\hat{#1}(#2)}

\def\n4sym{{\cal N}=4 SYM}
\def\>{\rangle}
\def\<{\langle}
\def\weight[#1,#2,#3]{\{(#1),#2,#3\}}
\def\ads[#1]{$\text{AdS}_{#1}$}

\newcommand{\be}{\begin{equation}}
	\newcommand{\ee}{\end{equation}}
\newcommand{\ba}{\begin{align}}
	\newcommand{\ea}{\end{align}}
\newcommand{\bs}{\begin{split}}
	\def\sess\end{split}
\newcommand{\vect}[1]{{\boldsymbol{#1}}}

\def \bea {\begin{eqnarray}}
	\def \eea {\end{eqnarray}}
\def \bea* {\begin{eqnarray*}}
	\def \eea* {\end{eqnarray*}}

\def \be {\begin{equation}}
	\def \ee {\end{equation}}
\def \bes {\begin{equation*}}
	\def \ees {\end{equation*}}

\title{Signatures of Bulk Black Hole Merger from Semi-classical 2d CFT}
\author[a]{Souvik Banerjee}
\emailAdd{souvik.banerjee@uni-wuerzburg.de}
\affiliation[a]{Institut für Theoretische Physik und Astrophysik, Julius-Maximilians-Universität Würzburg,\\ Am Hubland, 97074 Würzburg, Germany} 
\author[b,c]{and Gideon Vos}
\emailAdd{gideonvos@kias.re.kr}
\affiliation[b]{School of Physics, Korea Institute for Advanced Study, 85 Hoegi-ro, Dongdaemun-gu, Seoul 02455, Republic of Korea}
\affiliation[c]{Central European Institute for Cosmology, \\ FZU, Na Slovance 1999/2, 182 21 Prague 8, Czech Republic\\}

\date{}

\abstract{We study the real-time bulk AdS$_3$ two-into-two scattering amplitude of conical defects. This is done through means of a previously developed method of systematic approximation of the monodromy problem for the Virasoro blocks of the four-point HHHH correlator in 2d CFT in the large central charge regime. We find that dialing the external scaling dimension triggers a transition from a scattering phase to an intermediate black hole phase. This transition occurs before the individual heavy operators exceed the BTZ mass gap.}
\begin{document}
	\maketitle

	%\tableofcontents

	\section{Introduction}
	Semi-classical two-dimensional conformal field theory with large central charge is generally expected to be dual to semi-classical gravity in a three-dimensional asymptotically Anti-de Sitter spacetime. While asymptotically AdS$_3$ gravity is an enormous simplification over higher-dimensional gravitational models it is still sophisticated enough to contain complex phenomena. Phenomena that among others include multi-centered solutions, black hole collapse and black hole mergers. 
	
	There have been many earlier works in AdS/CFT that explored bulk gravity in the probe regime, where particles are released from the boundary that do not backreact with the geometry. \cite{Asplund:2014coa, Fitzpatrick:2014vua, Fitzpatrick:2015zha, Chen:2016kyz, Chen:2016dfb, Balasubramanian:2017fan}. Usually these probe particles can be realized in the CFT dual in the generalized free theory regime or in heavy-light perturbation theory. It is a parametrically harder problem to study the gravitational dynamics of non-stationary backreacting bulk geometries that satisfy Einstein's equations. The dual CFT computation would involve out-of-equilibrium physics in strongly coupled field theory. Though difficult it is possible to construct general relations between the bulk geometries and boundary theory states \cite{Hulik:2016ifr,Raeymaekers:2022sbu, Vos:2018vwv, Anous:2016kss}. Apart from \cite{Anous:2016kss} these approaches are in a sense `non-constructive', they relate a mathematical problem in the boundary to a mathematical problem in the bulk. Without a solution we do not see `in-situ' the manifestation of bulk gravitational physics in the CFT. 
	
	In this paper we attempt to provide such an analytical example of the manifestation of black hole collapse in 2d CFT. We will compute the overlap of a two heavy operator state onto itself after Lorentzian time-evolution. This corresponds to the semi-classical scattering amplitude of bulk 2-into-2 scattering. This is an interesting scattering experiment as in principle it should have two very distinct modes depending on the mass of the initial particles. Even at zero angular momentum there exists a mass gap in the spectrum of the BTZ black hole \cite{Banados:1992wn}. The mass of the lightest BTZ black hole is parametrically separated from the mass of global empty AdS by
	\begin{equation}
		E_{\text{min BTZ}} - E_{\text{empty AdS}} = \frac{1}{8}\frac{R_{\text{AdS}}}{G_N},
	\end{equation}
	where $R_{\text{AdS}}$ is the AdS radius and $G_N$ is Newton's constant. When the energy of the initial state is too low a transition to an intermediate black hole state is classically forbidden. On the other hand, if we cross a transition point the intermediate states are dominated by black hole formation along the lines of the process described in \cite{Matschull:1998rv}\footnote{We would like to thank A. Mukhopadhyay for pointing out we had neglected to mention this work.}.  Despite that the transition to an intermediate black hole phase is not exclusively tied to the initial energy of the state. Due to the non-trivial Brown-Henneaux boundary charges \cite{Brown:1986nw}, some of the energy of the initial state is potentially locked up in these conserved charges and cannot participate in the formation of a black hole. Resolving this problem is in general a complicated procedure that involves constructing the restriction of the stress tensor expectation value to a specific time-slice and identifying to which Virasoro coadjoint orbit the resulting coadjoint vector belongs \cite{Banerjee:2018tut,Vos:2018vwv}
	
	Therefore, since kinematic arguments alone are not sufficient, we will compute the conformal blocks that enter transition amplitude. We will analytically continue the spacetime coordinates to the Lorentzian regime. Subsequently, we will see whether there is a quantitative shift in the behavior of the conformal blocks as a function of Lorentzian time $t$ as the scaling dimension of the external primary operators is continuously dialed.
	
	\subsection{Methodology}
	
\begin{figure}
		\centering
		\includegraphics[scale=0.7]{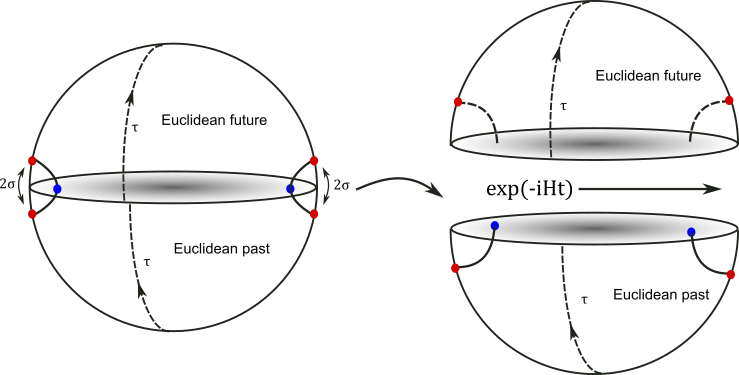}
		\caption{Our set-up, we will prepare both the CFT in-state and out-state by Euclidean path integral in radial quantization. In the middle we will evolve the initial in-state with the Lorentzian time-evolution operator before projecting it back on the out-state. In the bulk we will assume our state corresponds to some initial two-centered solution. }
		\label{2in2LorentzianEvolution}
	\end{figure}

\begin{figure}
		\centering
		\includegraphics[width=0.5\linewidth]{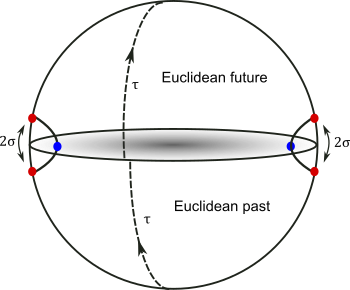}
		\caption{The Euclidean path integral that prepares both the in-state and the out-state. At Lorentzian time $t=0$ we want to maintain the interpretation of a scalar product of our initial state, hence the adjoint operator position on the upper hemisphere are mirrored with respect to the equation where our initial state lives. The parameter $\sigma$ controls the separation of the mirrored operators.}
		\label{AdSstatepreparation}
	\end{figure}

	We will construct our transition amplitude in the spirit of real-time gauge/gravity duality as proposed in \cite{Skenderis:2008dg, Skenderis:2008dh} and subsequently \cite{Marolf:2017kvq}. In order to construct the Lorentzian transition amplitude we will prepare our initial CFT state by Euclidean path integral, see fig. \ref{AdSstatepreparation}. The resulting state will be evolved through Lorentzian time and projected back onto the adjoint of the initial state which is also prepared by Euclidean path integral, see fig. \ref{2in2LorentzianEvolution}.
	
	At the level of the CFT, we will consider our state to be created through radial quantization and to live on the unit circle. As a result, we can create the adjoint state by reflecting the location of our operators around the unit circle. At Lorentzian time $t=0$ our transition amplitude hence takes on the form of a Euclidean four-point function.
	
	In compliance with the Brown-Henneaux central charge formula $c = \frac{3}{2} \frac{R_{\text{AdS}}}{G_N}$ we will assume the central charge to be a very large number in order to facilitate a semi-classical gravitational bulk. In addition, in order to create bulk objects capable of causing significant backreaction to the bulk geometry we will assume our CFT operators are `heavy', i.e. their scaling dimensions $H$ are commensurate with respect to the central charge $H\sim c$.
	
	All things considered, at Lorentzian time $t=0$ our problem corresponds to computing the CFT four-point function
	\begin{equation}
		\langle \mathcal{O}(-1-\sigma) \mathcal{O}(-1+\sigma)\mathcal{O}(1+\sigma)\mathcal{O}(1-\sigma)\rangle,
		\label{fourpointfunction}
	\end{equation}
	where we have already assumed the separation parameter $\sigma$ to be small. In the upcoming discussion we will mostly forego the separation parameter $\sigma$ in terms of the conformal cross-ratio $x$ The relationship between the two quantities is extremely simple though
	\begin{equation}
		x = \sigma^2.
	\end{equation}
	In order to establish the analytical continuation to Lorentzian time it will be helpful to write the mixed Euclidean-Lorentzian time evolution in terms of the Schr\"odinger picture
	\begin{equation}
		\langle \mathcal{O}\mathcal{O} e^{-H (1+\sigma -1)} \,|e^{-iHt}\, |\, e^{-H(1-1+\sigma)}\mathcal{O}\mathcal{O}\rangle = \langle \mathcal{O}\mathcal{O}|\, e^{-H(2\sigma +i t)}\,|\mathcal{O}\mathcal{O}\rangle,
	\end{equation}
	from this we can read off an important intermediate conclusion. We can compute the overlap of the initial state after Lorentzian time-evolution with its adjoint at time $t=0$ simply by analytically continuing the separation parameter $\sigma$ to the upper half-plane. This will be our strategy, we will approximate the conformal blocks of the four-point function with real $x$ and subsequently analytically continue $x$ to the upper half-plane.
	
	\subsection{Results and overview}
	We have two parameters to tune, the central charge $c$ and the real separation parameter $\sigma$ (or equivalently the cross-ratio $x$). As we will discuss in the next section where we will build some intuition for the bulk geometry dual to our initial state, there is a hierarchy to the order of magnitude of our parameters, we will insist that
	\begin{equation}
		1/c \ll \sigma \ll 1.
	\end{equation}
	This will ensure that the energy available at scattering is below inverse Planck length. The goal will be to approximate the conformal blocks of the four-point function \eqref{fourpointfunction} in the OPE channel
	\begin{equation}
		\langle \wick{ \c{\mathcal{O}} (-1-\sigma) \c{\mathcal{O}} (-1+\sigma)}\wick{\c{\mathcal{O}}(1+\sigma)\c{\mathcal{O}}(1-\sigma)}\rangle,  
	\end{equation}
	since $\sigma$ is small it is tempting to work out the OPEs term by term. In this case we find that the resulting sum takes on the schematic form
	\begin{equation}
		\mathcal{F}(H,H_p, x) = x^{-2H+H_p}(1 + \frac{1}{2}H_p\, x + \# (Hx)^2 + \# (Hx)^3 + ...).
	\end{equation}
	While strictly speaking this series converges, if we assume heavy operators $H\sim c$ and/or heavy intermediate primaries $H_p \sim c$ we find that for our hierarchy of parameters the first few partial sums are very poor approximations for the conformal block. Instead we will apply the monodromy method to compute the exponent of the semi-classical conformal block $f(h,h_p,x)$ which at large $c$ is closely related to the conformal block
	\begin{equation}
		\mathcal{F}(H,H_p,x) = (1 + \mathcal{O}(1/c))e^{-\frac{c}{6}f(h,h_p,x) +\mathcal{O}(1/c)}.
		\label{conformalblock}
	\end{equation}
	Here the function $f(h,h_p,x)$ exclusively depends on the scaling dimensions through the order unity ratios
	\begin{equation}
		h = \frac{6H}{c}, \hspace{10mm} \text{and} \hspace{10mm} h_p = \frac{6H_p}{c}.
	\end{equation}
	We will construct the semi-classical block through means of the monodromy method. As an intermediate step this will require solving a Fuchsian linear ODE with four regular singular points. We will approximate the solutions by applying a method of dominant balances proposed in \cite{Hadasz:2006rb}.
	
\begin{figure}
		\centering
		\includegraphics[width=1\linewidth]{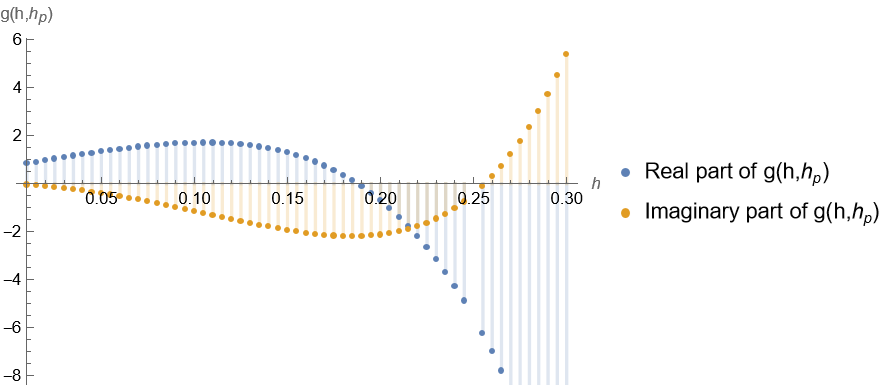}
		\caption{A numerical plot of $g(h,h_p)$ as a function of $h$ with fixed $h_p = \frac{1}{50}$. Note the sign flip of the real part which occurs well before the black hole threshold of the initial particles at $h=1/4$.}
		\label{TracePlot1}
	\end{figure}

	To leading order in $x$ we once again reproduce
	\begin{equation}
		e^{-\frac{c}{6}f(h,h_p,x)} \approx x^{-2H + H_p}, 
	\end{equation}
	including the first subleading correction we find that the leading order contribution receives a multiplicative dressing factor
	\begin{equation}
		e^{-\frac{c}{6}f(h,h_p,x)} = x^{-2H + H_p} \times e^{ \frac{\alpha_p (2H+H_p)g(h,h_p)}{4\pi \sin(\pi \alpha_p)}x \; +\, \text{subl.}},
		\label{dressedsemiclassicalblock}
	\end{equation}
	where $\alpha_p$ is given by
	\begin{equation}
		\alpha_p = \sqrt{1-4h_p/c}.
	\end{equation}
	The function $g(h,h_p)$ that appears in the exponent of the dressing factor in \eqref{firstorderdressedblock} is related to the trace of a monodromy matrix that we will encounter in section \ref{sec:firstcorrection}, it is generically complex-valued. Note that the rest of the factors of the exponent are manifestly real and positive, as a consequence after analytic continuation
	\begin{equation}
		x = \sigma^2 \Longrightarrow (\sigma + it)^2 = x + 2i\sigma t - t^2,
	\end{equation}
	the qualitative behavior of the semi-classical block as a function of time $t$ is determined by the phase of $g(h,h_p)$. Determining $g(h,h_p)$ analytically is a very difficult task, but it can be computed numerically, see fig. \ref{TracePlot1} for a plot of $g(h,h_p)$ as a function of $h$ with fixed $h_p$. It can be seen that there is sign flip in the real part that occurs before the point $h=1/4$, i.e. $H=c/24$. This sign flip indicates a transition from a growing oscillating phase to an exponentially decreasing phase as a function of $t$. 
	
	At $H=c/24$ the individual initial bulk objects have exceeded the BTZ mass threshold and are black holes themselves. Interestingly, below the threshold we already witness a transition to an exponentially decaying phase. Prior to the decaying phase we find a growing oscillating phase which is consistent with objects that either do not interact or elastically scatter. If on the other hand the two bulk constituents collapse to an intermediate black hole state then the amplitude for the black hole to decay by Hawking emitting the exact two constituents it was build from is allowed but highly suppressed.
	Therefore we interpret the transition to a decaying phase as a signature of bulk black hole collapse.
	
\begin{figure}
		\centering
		\includegraphics[width=1\linewidth]{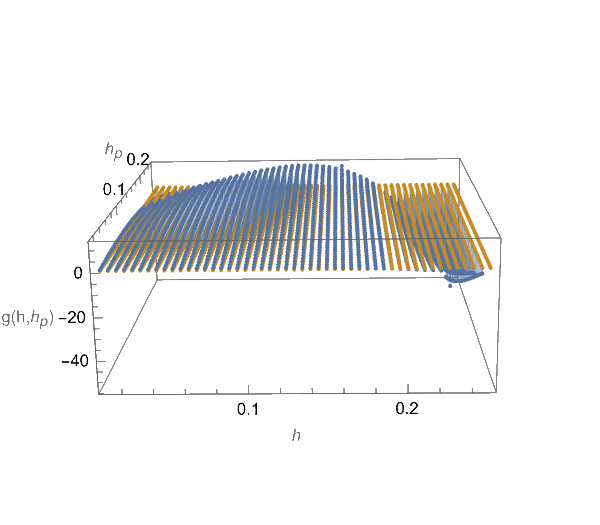}
		\caption{A numerical 3d plot of the real part $g(h,h_p)$ as a function of $h$ and $h_p$. The orange plane is the zero plane. Note the sign flip of the real part appears to be independent of $h_p$, suggestive that the transition is a property of the CFT state and not the intermediate primary.}
		\label{TracePlot3d}
	\end{figure}

	Indicative that the transition is a property of the CFT state and not the intermediate primary of the conformal block, we can see in fig. \ref{TracePlot3d} that the transition point appears to be independent of $H_p$, i.e. the choice of intermediate primary in the OPE channel.
	
	The remainder of this paper is divided as following: in the next section, section \ref{sec:kinematics} we will attempt to gain some insight into the bulk geometry by glueing a Ba\~nados geometry to our initial slice through means of the stress tensor expectation value. In section \ref{sec:monodromyreview} we will briefly review the monodromy method. In section \ref{sec:fourheavyoperators} we will apply the method of dominant balances of \cite{Hadasz:2006rb} to construct the leading part of the semi-classical conformal block. In section \ref{sec:firstcorrection} we will construct the first-subleading correction to the conformal block. We will close off with a discussion section. For those who might find it useful we have added an appendix where we reduce our bases of leading solutions to Legendre functions.

	\section{AdS kinematics}
	\label{sec:kinematics}
	To set the physical intuition and establish the motivation of our method it will be useful to delve into some AdS kinematics. We want to consider the Lorentzian time-evolution of a state prepared in the bulk by Euclidean path integral. As was anticipated in the classic work \cite{Verlinde:1989ua} the wave functionals of 3d AdS gravity are given by the conformal blocks. In the context of the AdS/CFT correspondence this was discussed in the works \cite{Skenderis:2008dg, Skenderis:2008dh} and \cite{Marolf:2017kvq}, the role with the identity block problem specifically was clarified in \cite{Raeymaekers:2022sbu} which very directly linked the conformal block problem to an initial slice geometry of a multi-centered solution in AdS.
	
	In all these cases the initial Lorentzian state is prepared by sewing a Euclidean cap to the initial Cauchy slice and performing a Euclidean path-integral on the on the cap. From the boundary theory perspective we will consider the boundary state obtained by path-integrating the boundary disk of the cap in radial quantization. Similarly the out-state is created by path-integrating along a Euclidean cap attached to the future Cauchy slice. 
	
	As discussed in  in \cite{Marolf:2017kvq, Raeymaekers:2022sbu}, we can prepare Wilson lines in the Euclidean bulk by acting with local primary CFT operators on the boundary of the cap at the endpoints of the Wilson lines\footnote{Disclosure: the analysis of \cite{Marolf:2017kvq} is based on the linearized regime, we will consider backreacting heavy operators which will be well outside the linearized regime.}. 
	
	The geometry on the Cauchy slice will presumably be some kind of initial slice of a fully backreacted two-centered solution and hence feasibly very complicated. We will remain mostly agnostic about the details of the initial slice, but by applying these dualities we can at least gain some intuition without delving into detailed bulk reconstruction. By considering fig. \ref{AdSstatepreparation} we can see that our set-up produces an initial slice geometry of a multi-centered solution with two centers. We can also see that heuristically $\sigma$ parametrizes how close to the boundary the two centers are.

\begin{figure}
		\centering
		\includegraphics[width=0.4\linewidth]{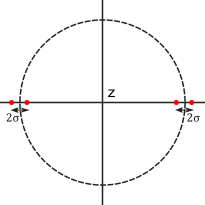}
		\caption{The distribution of mirror pairs of operators that prepares the relevant in- and out-state in the dual CFT on the radial plane.}
		\label{mirrorpairs}
	\end{figure}

	The dual CFT state is prepared by radial quantization on the disk that forms the boundary of the cap. We will insert two Hermitian scalar CFT primary operators on the radial disk located at respectively $1-\sigma$ and $-1+\sigma$. We will want our out-state to be adjoint vector of the initial state at Lorentzian time $t=0$ so we will sew another Euclidean disk onto our original disk with operators at $1/(1-\sigma) \sim 1+\sigma$ and $1/(-1+\sigma) \sim ~ -1 - \sigma$. See fig. \ref{mirrorpairs}, we will assume throughout that $\sigma \ll 1$ and use the linearized expression for the adjoint operator locations. Using the full expression instead will modify the conformal cross-ratio $x$ by an order $\mathcal{O}(\sigma^3)$ shift which will be outside of our order of approximation. Notice that this choice of reflection implicitly places our initial Lorentzian state on the unit circle.
	
	Note that this is not an eigenstate of the Hamiltonian, we will take our operators to be heavy, i.e. there scaling dimensions $H$ will be of the same order as the central charge $c$ which we assume to be large. As a result our initial state will potentially be a very complex high-energy out-of-equilibrium state.
	
	We can compute the dominant part of the stress tensor expectation value on the CFT Lorentzian initial state. This can by accomplished by transforming the operator $\hat{T}(z)$ to the cylinder by undoing the radial map $z= \exp(2\pi i x/L_{\text{CFT}})$
	\begin{align}
		& \hat{T}(x) = \left(\frac{dz}{dz}\right)^2 \hat{T}(z(x)) + \frac{c}{12}S[z,x] \nonumber \\
		& = -\frac{4\pi^2}{L_{\text{CFT}}^2} \hat{T}(e^{2\pi i x/L_{\text{CFT}}}) + \frac{\pi^2 c}{6L_{\text{CFT}}^2}
		\label{cylindertransformation}
	\end{align}
	We can utilize this expression to derive the stress tensor expectation value on the cylinder. First notice that on the radial plane the Virasoro Ward identity dictates
	\begin{equation}
		\langle \hat{T}(z)\mathcal{O}_H(Z_1)...\mathcal{O}_H(z_4)\rangle = \sum_{i=1}^{4} \left(\frac{H}{(z-z_i)^2} + \frac{\partial_{z_i}}{z-z_i}\right)\langle \mathcal{O}_H(z_1)...\mathcal{O}_H(z_4)\rangle.
	\end{equation}
	As we will see in the upcoming section, in the regime of large central charge and small $\sigma$ our correlator to leading order factorizes
	\begin{equation}
		\langle \mathcal{O}_H(1+\sigma)\mathcal{O}_H(1-\sigma)\mathcal{O}_H(-1+\sigma)\mathcal{O}_H(-1-\sigma)\rangle \sim (z_1-z_2)^{-2H}(z_3-z_4)^{-2H} = (2\sigma)^{-4H}
	\end{equation}
	If we define the normalized expectation value $T(z)$
	\begin{equation}
		T(z) = \frac{\langle \hat{T}(z)\mathcal{O}_H(z_1)...\mathcal{O}_H(z_4)\rangle}{\langle \mathcal{O}_H(z_1)...\mathcal{O}_H(z_4)\rangle}.
	\end{equation}
	Then we find that the pole structure of $T(z)$ up to subleading corrections in $\sigma$ is given by
	\begin{equation}
		T(z) =\left(\sum_{\{\pm,\pm\}} \frac{H}{(z\pm 1 \pm \sigma)^2}\right) + \frac{4H(\sigma^2-1-z^2)}{(z-1-\sigma)(z+1-\sigma)(z-1+\sigma)(z+1+\sigma)}. 
	\end{equation}
	We can bring this to the cylinder expression through means of \eqref{cylindertransformation}. We will use the cylinder parametrized stress-energy parameters as sources for the Ba\~nados geometry. 
	
	\subsection{Ba\~nados geometries and the quasi-local stress tensor.}
	To obtain some intuition for the initial slice geometry prepared by our Euclidean path integral we will consider the ADM mass of our bulk geometry. Take the gravitational sector of the bulk to be given by the semi-classical action
	\begin{equation}
		S[g_{\mu\nu}] = -\frac{1}{16\pi G_N}\int_{\mathcal{M}} d^3x \, \sqrt{-g}\left(R-\frac{6}{R_{\text{AdS}}^2}\right) - \frac{1}{8\pi G_N}\int_{\partial\mathcal{M}} d^2 x \sqrt{-\gamma} \Theta + \frac{1}{8 \pi G_N R_{\text{AdS}}}\int_{\partial\mathcal{M}} d^2 x \, \sqrt{-\gamma}. 
	\end{equation}
	Here $\mathcal{M}$ is the solid Lorentzian cylinder, the first term is the Einstein-Hilbert action with negative cosmological constant. The second term is the Gibbons-Hawking-York boundary term that renders the action finite on-shell, here $\Theta$ is the trace of the extrinsic curvature of the boundary surface $\partial\mathcal{M}$ as embedded in $\mathcal{M}$. The third term is a counter term that subtracts the divergences from the quasi-local stress tensor. 
	
	The most general metric that solves the equation of motion with asymptotically AdS boundary conditions is known and is given by the Ba\~nados metric \cite{Banados:1998gg}. In Fefferman-Graham coordinates (convention taken from \cite{Sheikh-Jabbari:2016unm}) it takes the form
	\begin{equation}
		ds^2 = \frac{R_{\text{AdS}}}{r^2}dr^2 - \left(\frac{r^2}{R_{\text{AdS}}^2} - \frac{R_{\text{AdS}}^2}{r^2}L(x)\bar{L}(\bar{x})\right)dx d\bar{x} + L(x)dx^2 + \bar{L}(\bar{x}) d\bar{x}^2.
		\label{banados}
	\end{equation}
	%\begin{equation}
	%ds^2 = \frac{R^2}{r^2}dr^2 + 4GR \, L(x) dx^2  + 4GR\, \bar{L}(\bar{x}) d\bar{x}^2 + \left(R^4 r^2 + \frac{16 G^2}{R^2 r^2} L(x) \bar{L}(\bar{x})\right) dx d\bar{x},
	%\end{equation}
	Here $R_{\text{AdS}}$ is the AdS radius and $x$ and $\bar{x}$ are lightcone coordinates given by $x = t + R_{\text{AdS}}\phi$ and $\bar{x} = -t + R_{\text{AdS}}\phi$. The functions $L(x)$ and $\bar{L}(\bar{x})$ are periodic functions of their respective lightcone coordinate that parametrize the family of metrics.
	
	Following the work \cite{Balasubramanian:1999re}, by including the counter term we can define a well-defined quasi-local stress tensor defined at the boundary
	\begin{equation}
		T^{\mu\nu} = \frac{2}{\sqrt{-\gamma}}\frac{\delta S}{\delta \gamma_{\mu\nu}}= \frac{1}{8\pi G} \left( \Theta^{\mu\nu} - \Theta \gamma^{\mu\nu} + \frac{1}{R_{\text{AdS}}} \gamma^{\mu\nu}\right).
	\end{equation}
	Where the extrinsic curvature $\Theta^{\mu\nu}$ can be expressed in terms of the covariant derivative of a unit vector $\hat{\eta}^{\mu} = r/R_{\text{AdS}} \, \delta^{\mu,r}$ normal to the boundary 
	\begin{equation}
		\Theta^{\mu\nu} = -\frac{1}{2}\left(\nabla^{\mu} \hat{\eta}^{\nu} + \nabla^{\nu} \hat{\eta}^{\mu}\right).
	\end{equation}
	With this expression for the stress tensor we can express the mass of a slice of the metric \eqref{banados} as
	\begin{equation}
		M = \frac{1}{8\pi G R_{\text{AdS}}} \int_0^{2\pi} d\phi\; L(\phi) + \bar{L}(\phi).
	\end{equation}
	As a consequence, if we perform the following AdS/CFT dictionary identification
	\begin{equation}
		\frac{L(x)}{8\pi G R_{\text{AdS}}} = T(x)|_{\text{Im}(x)=0}, \hspace{12mm} \frac{L(x)}{8\pi G R_{\text{AdS}}} = \bar{T}(\bar{x})|_{\text{Im}(\bar{x})=0}.
		\label{adscftdictionary}
	\end{equation}
	and subsequently set $R_{AdS} = \frac{L_{\text{CFT}}}{2\pi} = 1$ and set the CFT central charge to the Brown-Henneaux value
	\begin{equation}
		c = \frac{3}{2}\frac{R_{AdS}}{G_N},
	\end{equation}
	we consistently reproduce the vacuum energy density of AdS$_3$ and the Bekenstein-Hawking temperature of the BTZ black hole\footnote{In order to do this, set $L$ and $\bar{L}$ to constant positive values and compare with the temperature of a heavy primary state with scaling dimension dictated by \eqref{adscftdictionary} as computed in \cite{Fitzpatrick:2015zha}.}. And in this case we find that the mass of the bulk geometry is given by
	\begin{equation}
		M = 2\frac{H}{\sigma} - \frac{1}{8 G_N}.
	\end{equation}
	This expression allows us to gain some bulk insight through means of a very heuristic back-of-the-envelop argument. In the exact same way we can confirm that the state created by a single scalar primary operator in the origin with scaling dimension $H$ is simply given by
	\begin{equation}
		M_{\text{primary}} = H.
	\end{equation}
	Since we assumed that $\sigma$ is a small number it seem reasonable to assume that the field excitations on the unit circle created by the radial path integral are close to local. As such the conventional AdS/CFT dictionary suggests that we can heuristically interpret the initial state in the bulk as two local objects close to the AdS boundary at opposite points with respect to the center of AdS. We find that the gravitational potential energy of a single object is
	\begin{equation}
		E_{\text{potential}} \approx \frac{H}{\sigma} - H = H\left(\frac{1}{\sigma} -1\right).
	\end{equation}
	As such we find that the gravitation potential energy is proportional to $1/\sigma$. The central is inversely proportional to $G_N$, $c \propto 1/G_N$. and the Plank length is proportional to $l_{\text{Planck}} \propto G_N$. So we find that in order to avoid collision energies of the order of the Planck scale we should maintain the following hierarchy of orders of magnitudes for our parameters
	\begin{equation}
		1/c \ll \sigma \ll 1.
	\end{equation}
	As mentioned in the introduction, this suggests that the monodromy method is a  well-suited tool to tackle the dual CFT problem. This method will be reviewed briefly in the upcoming section.
	
	%This gives us some insight as it suggests that in order to have a semi-classical bulk we  require the following regime of approximation $1/c \ll \sigma \ll 1$. This can be seen from some basic relativistic kinematics for a single conical defect take
	%\begin{equation}
	%-m^2 = p_{\mu}p^{\mu} = -E^2 + |p|^2.
	%\end{equation}
	%Take $E = H/\sigma$, $m=H$ and we find $|p|^2/m^2 \sim 1/\sigma^2$. Hence if $\sigma \sim 1/c$ we would be in the deep ultra-relativistic regime.  

	\section{Review of the monodromy method}
	\label{sec:monodromyreview}
	The monodromy method is a technique that leads to the construction of the conformal blocks in the semiclassical large-$c$ regime. The original concept was developed in \cite{BELAVIN1984333} and reviewed in \cite{Hartman:2013mia, Fitzpatrick:2014vua}. It attempts to exploit the Liouville theory-inspired conjectured exponentiation of heavy operators and the factorization of light operators. In effect, exponentiation states that both  correlators of heavy operators $\mathcal{O}$ and their constituent conformal blocks scale as
	\begin{equation}
		\langle \mathcal{O}_1(x_1)...\mathcal{O}_n(x_n)\rangle = e^{-\frac{c}{6} f(x_i, H_i)}, 
	\end{equation}
	where $H_i$ are the scaling dimensions of the heavy operators.
	Factorization states that the subsequent addition of light operators to the correlator $\phi_i(z_i)$ provides a factor contribution to the correlators and conformal blocks
	\begin{equation}
		\langle \phi_1(z_1)...\phi_m(z_m)\mathcal{O}_1(x_1)...\mathcal{O}_n(x_n)\rangle = \psi(z_i, h_i, x_i, H_i)e^{-\frac{c}{6}f(x_i,H_i)}.
	\end{equation}
	Here we have denoted, by $h_i$, the scaling dimensions of the light operators $\phi_i$.
	We will exploit this expression by constructing a specific light ghost primary with a null state at level two. Take this state to be $|\phi\rangle$, it satisfies
	\begin{equation}
		(L_{-2} + k L_{-1}^2)|\phi\rangle =0.
	\end{equation}
	We can find two constraint equations by acting respectively with $L_2$ and $L_{1}^2$ on the left-hand side
	\begin{align}
		& L_2(L_{-2} + k L_{-1}^2)|\phi\rangle =0 \longrightarrow (6k + 4)h_{\phi} +\frac{c}{2}=0,\\
		& L_{-1}^2(L_{-2} + k L_{-1}^2)|\phi\rangle =0 \longrightarrow 6h_{\phi} + 2k h_{\phi} (4h_{\phi}+2) =0.
	\end{align}
	We can use these two constraints to solve for both $h_{\phi}$ and $\alpha$, under the assumption that $c\gg 1$ these solutions simplify to
	\begin{align}
		& k = \frac{c}{6} + \mathcal{O}(c^0),\\
		& h_{\phi} = -\frac{1}{2} + \mathcal{O}(1/c).
	\end{align}
	Due to the latter of the two expressions the relevant primary state is a ghost state and we can confirm that it is a light operator. Take $\phi(z)$ to be the primary operator that generates the ghost state
	\begin{equation}
		\phi(0)|0\rangle = |\phi\rangle.
	\end{equation}
	Inserting the operator in a correlator consisting of heavy operators results in
	\begin{equation}
		\langle \mathcal{O}_1(x_1)...\mathcal{O}_n(x_n)\phi(z)\rangle = \psi(z) e^{-\frac{c}{6}f(x_i)}.
	\end{equation}
	In which case the null vector decoupling equation takes the form
	\begin{equation}
		\langle \mathcal{O}_1(x_1)...\mathcal{O}_n(x_n) \left([L_{-2},\phi(z)]+\frac{c}{6}[L_{-1},[L_{-1},\phi(z)]]\right)\rangle =0.
		\label{predecouplingequation}
	\end{equation}
	Handling both terms separately
	\begin{equation}
		\langle \mathcal{O}_1(x_1)...\mathcal{O}_n(x_n) [L_{-1},[L_{-1},\phi(z)]]\rangle = e^{-\frac{c}{6}f(x_i)}\frac{\partial^2 \psi}{\partial z^2},
	\end{equation}
	and
	\begin{align}
		& \langle \mathcal{O}_1(x_1)...\mathcal{O}_n(x_n) [L_{-2},\phi(z)]\rangle = \frac{1}{2\pi i} \oint dz' \frac{1}{z-z'} \langle T(z')\phi(z)\mathcal{O}_1...\mathcal{O}_n\rangle \nonumber \\
		& = \left(\sum_{i=1}^n \frac{H_i}{(z-x_i)^2}-\frac{\frac{c}{6}\partial_{x_i}f}{z'-x_i} +\mathcal{O}(c^{0})\right) \psi(z)e^{-\frac{c}{6}f}
	\end{align}
	We define the meromorphic function $T(z)$ as 
	\begin{equation}
		T(z) = \sum_{i=1}^n \frac{H_i}{(z-x_i)^2}-\frac{\frac{c}{6}c_i}{z'-x_i},
		\label{stresstensor}
	\end{equation}
	where we define the accessory parameters
	\begin{equation}
		c_i \equiv \frac{\partial f}{\partial x_i}.
	\end{equation}
	In the regime $c\gg 1$ the expression \eqref{predecouplingequation} takes the form
	\begin{equation}
		\psi''(z) + \frac{6}{c}T(z)\psi(z) = 0.
		\label{Fuchsequation}
	\end{equation}
	Due to the meromorphic nature of $T(z)$ this is an equation of Fuchsian class. The goal will be to obtain a set of constraints on the accessory parameters. This is the approach of the monodromy method, which can be summarized in the following way:
	\begin{itemize}
		\item Solve the linear ODE for undetermined accessory parameters.
		\item Predetermine the class of the monodromy of the solutions for a loop around two operator insertions based on the OPE channel of the conformal block.
		\item Compute the monodromy of the solutions with undetermined accessory parameters around the same loop.
		\item Equate the monodromy matrices of the last two bullet points up to unitary conjugation and use them as constraints to solve for the accessory parameters.
		\item Integrate the resulting accessory parameters to obtain the semi-classical conformal block.
	\end{itemize}
	The first bullet point is by far the hardest step and can only be performed analytically in very select cases. The majority of section \ref{sec:fourheavyoperators} will be devoted towards obtaining systematic corrections to the monodromy matrix in terms of the cross ratio $x$.

	\subsection{A note on regularity conditions}
	The accessory parameters are not entirely independent since we will need to specify the behavior of $T(z)$ at infinity. There are two possible cases, either we demand regularity at infinity, or we demand that infinity is also a regular singular point in which case we need to prescribe the strength of the singular point at infinity.
	
	In either of the two cases this can be derived from the transformation rule of the stress tensor under conformal transformations $w(z)$
	\begin{equation}
		T'(w) = \left(\frac{\partial w}{\partial z}\right)^{-2} T(z) +\frac{c}{12}S[z,w],
	\end{equation}
	with Schwarzian derivative
	\begin{equation}
		S[f,z] = \frac{f'''}{f'} - \frac{3}{2}\left(\frac{f''}{f'}\right)^{2}.
	\end{equation}
	As a consequence, under an inversion $w=1/z$ we find
	\begin{equation}
		T(z) = \frac{1}{z^4}T(1/z),
	\end{equation}
	With the explicit form of $T(z)$ given in \eqref{stresstensor} we can expand $T(1/z)$ around $z\rightarrow 0$ leading to
	\begin{equation}
		\frac{1}{z^4}T(1/z) \overset{z\rightarrow 0}{=}\frac{\sum_i c_i}{z^3} + \frac{\sum_i \frac{6 H_i}{c} - c_i x_i}{z^2} + \frac{\sum_i \frac{12 H_i}{c}x_i - c_ix_i^2}{z} + \mathcal{O}(z^0).
	\end{equation}
	Hence we can directly read off a set of constraints from the requirement that $T(z)$ does not blow up at infinity
	\begin{align}
		& \sum_{i=1}^n c_i = 0, \\
		& \sum_{i=1}^n c_i x_i - \frac{6H_i}{c} = 0,\\
		& \sum_{i=1}^n c_i x_i^2 -\frac{12 H_i}{c}x_i =0.
	\end{align}
	In the case that we do want infinity to be a regular singular point we can derive a (smaller) set of constraints. In this case the point at infinity is allowed to blow up as $z^2$, but we demand that the coefficient with which it blows up is fixed by a `scaling weight at infinity'. In this case we obtain the two constraints 
	\begin{align}
		& \sum_{i=1}^n c_i = 0, \\
		& \sum_{i=1}^n c_i x_i - \frac{6H_i}{c} = \frac{6 H_{\infty}}{c}.
	\end{align}

	\subsection{monodromy constraints}
	We need an additional $n-3$ constraints to solve for all accessory parameters, we will consider the monodromy group of the solutions to the Fuchsian equation \eqref{Fuchsequation}. 
	
	The monodromy of a loop around a single point is very easy to compute, in this case we can contract the loop to arbitrarily close around the singular point where the Fuchsian equation can be approximated by
	\begin{equation}
		\psi_{x_i}''(z) + \frac{6}{c}\frac{H_i}{(z-x_i)^2}\psi_{x_i}(z) = 0,
		\label{indicialequation}
	\end{equation}
	this is the indicial equation associated to the regular singular, it is trivially solved and has the two solutions
	\begin{equation}
		\psi_{x_i}^{\pm}(z) = (z-x_i)^{\frac{1}{2}\pm \frac{1}{2}\sqrt{1-24H_i/c}}.
	\end{equation}

	From this we can read of the monodromy matrix
	\begin{equation}
		\begin{pmatrix}
			\psi^+_{x_i}\\
			\psi^-_{x_i}
		\end{pmatrix}
		\rightarrow
		\begin{pmatrix}
			-e^{i\pi \alpha_i} & 0\\
			0 & -e^{-i\pi \alpha_i}
		\end{pmatrix}
		\begin{pmatrix}
			\psi^+_{x_i}\\
			\psi^-_{x_i}
		\end{pmatrix},
	\end{equation}
	where we define
	\begin{equation}
		\alpha_i = \sqrt{1-24H_i/c}.
		\label{characteristicmonodromy}
	\end{equation}
	We can project out the conformal blocks from the correlator through the structure of the OPE channel. Demanding that a loop around two contracted operator yields a monodromy matrix that is up to unitary transformation equivalent to  monodromy matrix around a single operator as given in \eqref{characteristicmonodromy}, where $H_i$ is set to the scaling dimension of the exchange primary. 
	
	We can use this as a constraint to fix the remaining accessory parameters. Subsequently integrating all of them leads to an expression for the conformal block.

	\subsection{A transformation rule for the accessory parameters}
	It will be useful to go to a different global conformal frame with canonical locations for the singular points $0, x, 1, \infty$. The accessory parameters are not invariant under these conformal transformations. Consider the transformation rule of $T(z)$, which, as long as $w(z)$ is fractional linear, is homogeneous.
	\begin{equation}
		T'(w) = \left(\frac{\partial w}{\partial z}\right)^{-2} T(z).
		\label{homogeneoustrafo}
	\end{equation}
	If we define $w_i = w(z_i)$ then we expect the right-hand side to have second- and first-order poles at $w_i$. We can isolate the transformed accessory parameters by computing the residue of the first-order pole of the transformed stress tensor
	\begin{equation}
		\tilde{c}_i = -\lim_{w\rightarrow w_i} \frac{\partial}{\partial_{w}}\, (w-w_i)^2 T(w).
	\end{equation}
	By substituting the right-hand side of \eqref{homogeneoustrafo} for $T(w)$ we can obtain a transformation rule for the accessory parameter
	\begin{equation}
		\tilde{c}_i = \left(\frac{\partial z}{\partial w}|_{w=w_i}\right) c_i - \frac{6}{c}H_i \frac{\partial}{\partial w_i} \log\left(\frac{\partial z}{\partial w}|_{w=w_i}\right).
		\label{accessorytransformation}
	\end{equation}
	This exact formula can be found in \cite{Anous:2019yku}, another detailed derivation can be found in \cite{Vos:2020clx}.

	\section{Four heavy operators}
	\label{sec:fourheavyoperators}
	Having reviewed the method we will utilize we turn our attention back to the heavy four point function. As discussed in the introduction this is the CFT problem that will give us insight into our bulk scattering process
	\begin{equation}
		\langle \mathcal{O}(-1-\sigma)\mathcal{O}(1-\sigma) \mathcal{O}(-1+\sigma)\mathcal{O}(1+\sigma)\rangle.
	\end{equation}
	where we take the separation $\sigma$ to be very small but finite. Through means of the (unnormalized) global conformal transformation
	\begin{equation}
		z(x) = -\sigma \frac{x-1-\sigma}{x+1+\sigma},
	\end{equation}
	this is mapped to a four-point function of conventional form
	\begin{equation}
		\langle \mathcal{O}(0)\mathcal{O}(x)\mathcal{O}(1)\mathcal{O}(\infty)\rangle
	\end{equation}
	where the conformal cross-ratio $x$ is simply given by
	\begin{equation}
		x=\sigma^2.
	\end{equation}
	Hence we see that as a consequence of the assumption that $\sigma$ is small that the regular singular points of our Fuchsian ODE will be very close to each other. We again emphasize that $\sigma$ is small but \textit{finite} as we will not be taking the confluent limit. 
	
	\begin{table}[]
		\centering
		\begin{tabular}{|c|c|c|}
			\hline
			$\tilde{c}_i$ & $x$ & $z$ \\
			\hline
			$\tilde{c}_1$ & $1 + \sigma$ & 0\\
			$\tilde{c}_2$ & $1 - \sigma$ & $x=\sigma^2$\\
			$\tilde{c}_3$ & $-1+\sigma$ & 1\\
			$\tilde{c}_4$ & $-1-\sigma$ & $\infty$\\
			\hline
		\end{tabular}
		\caption{A synopsis of the labeling and location of the singular points in both $x$- and $z-$coordinates.}
		\label{zxtable}
	\end{table}
	
	To keep things straight, the relevant labeling of the regular singular points is summarized in \ref{zxtable}. We can construct the relevant transformation rule for the accessory parameters by applying \eqref{accessorytransformation}, this results in
	\begin{equation}
		\tilde{c}_i = \frac{-2\sigma (1+\sigma)}{(z_i +\sigma)^2}c_i +\frac{2h}{z_i+\sigma},
		\label{ouraccessorytransformation}
	\end{equation}
	where $h$ is defined simply as
	\begin{equation}
		h= \frac{6 H}{c}.
	\end{equation}
	The stress tensor expectation value in $z$-coordinates is given by
	\begin{equation}
		T(z) = \frac{h}{z^2} + \frac{h}{(z-x)^2} + \frac{h}{(z-1)^2} - \frac{\tilde{c}_1}{z} - \frac{\tilde{c}_2}{z-x} - \frac{\tilde{c}_3}{z-1}.
	\end{equation}
	With constraints given by
	\begin{align}
		& \tilde{c}_1+\tilde{c}_2 + \tilde{c}_3 = 0,\\
		& x \tilde{c}_2 + \tilde{c}_3 = 4h.
	\end{align}
	We will now exploit the relative closeness of the singular points at 0 and $x$ to construct and approximation scheme for the solutions to the Fuchsian equation.
	
	\subsection{Approximating the solutions}
	We will take $\sigma \ll 1$, as a result the points $0$ and $x=\sigma^2$ are finitely separated but close to each other on the complex plane. We will apply the method proposed in \cite{Hadasz:2006rb} to split the stress tensor expectation value into two parts\footnote{see also \cite{Kusuki:2018nms} for a WKB-based analysis.}. First, we will use the first constraint to set $\tilde{c}_2 = - \tilde{c}_1 - \tilde{c}_3$ which allows us to separate $T(z)$ as
	\begin{equation}
		T(z) = T_0(z) - V(z),
	\end{equation}
	where
	\begin{equation}
		T_0(z) = \frac{h}{z^2} + \frac{h}{(z-x)^2} - \frac{x \tilde{c}_2}{z(z-x)},
		\label{leadingstresstensor}
	\end{equation}
	and
	\begin{equation}
		V(z) = - \frac{h}{(z-1)^2} +  \frac{\tilde{c}_3}{z(z-1)}.
		\label{correctionpotential}
	\end{equation}
	In the vicinity of $z=0$ (and by extension $z=x$) we have split up $T(z)$ into a part $T_0(z)$ that grow quadratically and a part $V(z)$ that contain all the terms that grow at most as a single-order pole. In order to fully justify this we should also establish the relative magnitude of the accessory parameters, we find from \eqref{ouraccessorytransformation} that
	\begin{align}
		& \tilde{c}_1 \propto \frac{c_1}{\sigma}, \label{transformedc1}\\
		& \tilde{c}_2 \propto \frac{c_2}{\sigma},\\
		& \tilde{c}_3 \propto \sigma c_3 + 2h,\\
		& \tilde{c}_4 = 0 \label{transformedc4}.
	\end{align}
	As we saw in section 2, to leading order all accessory parameters $c_i$ are to leading order proportional to $1/\sigma$. As such in the new conformal frame there is a hierarchy of relative magnitudes of the transformed accessory parameters $\tilde{c}_i$ in the new frame. This justifies the statement that the leading order contribution to the Fuchsian equation is given by
	\begin{equation}
		\rho''(z) + T_0(z) \rho(z) = 0.
		\label{leadingorderFuchsian}
	\end{equation}
	This is an ODE with just three regular singular points and, as a consequence, solvable. We will for now consider $\tilde{c}_2$ to be a free parameter, as a consequence the weight of the regular singular point at infinity is a function of $\tilde{c}_2$
	\begin{equation}
		h_{\infty} = 2h-x \tilde{c}_2.
		\label{weigthatinfinity}
	\end{equation}
	We can express the solutions to \eqref{leadingorderFuchsian} in terms hypergeometric functions. In which case it will be useful to start with the hypergeometric equation in Papperitz form
	\begin{align}
		&0= \rho''(z) +\left(\frac{1-\alpha-\alpha'}{z-a} + \frac{1-\beta-\beta'}{z-b}+\frac{1-\gamma-\gamma'}{z-c}\right)\rho' \nonumber \\
		& + \left(\frac{\alpha\alpha'(a-b)(a-c)}{z-a} + \frac{\beta\beta'(b-c)(b-a)}{z-b} + \frac{\gamma\gamma'(c-a)(c-b)}{z-c}\right)\frac{\rho}{(z-a)(z-b)(z-c)} .
	\end{align}
	The relevant constraints for the Papperitz coefficients are given by
	\begin{align}
		& \alpha \alpha' = h, \;\;\;\;\;\;\;\;\;\;\;\;\;\;\;\;\; \alpha + \alpha' =1 \\
		& \beta \beta' = h, \;\;\;\;\;\;\;\;\;\;\;\;\;\;\;\;\; \beta + \beta' =1 \\
		& \gamma \gamma' = 2h-x \tilde{c}_2, %\;\;\;\;\; \gamma + \gamma' =1
	\end{align}
	The coefficients, $\alpha$, $\alpha'$, $\beta$ and $\beta'$ are easily solved to give
	\begin{align}
		& \alpha = \frac{1}{2} + \frac{1}{2}\sqrt{1-4h}, \;\;\;\;\;\;\;\;\;\;\;\;\;\; \alpha' = \frac{1}{2} - \frac{1}{2}\sqrt{1-4h}, \\
		& \beta = \frac{1}{2} + \frac{1}{2}\sqrt{1-4h}, \;\;\;\;\;\;\;\;\;\;\;\;\;\; \beta' = \frac{1}{2} - \frac{1}{2}\sqrt{1-4h}. \\
		%&\gamma = \frac{1}{2} + \frac{1}{2}\sqrt{1+4h+4cx}, \;\;\;\;\; \gamma' = \frac{1}{2} - \frac{1}{2}\sqrt{1+4h+4cx}
	\end{align}
	In principle one of the $\gamma$ coefficients is free, but we will later on use regularity as a constraint to solve for the remaining coefficient.
	
\begin{figure}
		\centering
		\includegraphics[width=0.4\linewidth]{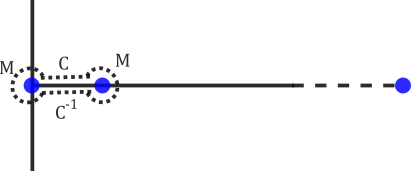}
		\caption{The contour along which the monodromy is computed. $C$ represents the basis transformation matrix that transforms the basis that has a diagonal monodromy matrix around $z=x$ to the basis that has diagonal monodromy matrix around $z=0$.}
		\label{monodromycontour}
	\end{figure}

	Formally a particular solution to the Papperitz equation is given by the Riemann P-symbol
	\begin{equation}
		\rho(z)=P\left\{ 
		\begin{matrix}
			0&1&\infty&\\
			\alpha&\beta&\gamma&\frac{z}{x}\\
			\alpha' &\beta'&\gamma'&\\
		\end{matrix}
		\right\}
	\end{equation}
	By using the identity
	\begin{equation}
		P\left\{ 
		\begin{matrix}
			x&y&\infty&\\
			\alpha&\beta&\gamma&z\\
			\alpha' &\beta'&\gamma'&\\
		\end{matrix}
		\right\}
		= (z-x)^{-\lambda}
		P\left\{ 
		\begin{matrix}
			x&y&\infty&\\
			\alpha+\lambda&\beta&\gamma -\lambda &z\\
			\alpha' +\lambda  &\beta'&\gamma'-\lambda &\\
		\end{matrix}
		\right\}
		\label{homographictrafo}
	\end{equation}
	and the identity
	\begin{equation}
		P\left\{ 
		\begin{matrix}
			0&1&\infty&\\
			0&0&a&z\\
			1-c &c-a-b&b&\\
		\end{matrix}
		\right\}
		= \,_2F_1(a,b;c,z)
	\end{equation}
	By using the homographic transformation \eqref{homographictrafo} we obtain a particular solution that is regular at the origin
	\begin{equation}
		\rho(z) = \left(\frac{z}{x}\right)^{\frac{1+\alpha}{2}}\left(\frac{z}{x}-1\right)^{\frac{1+\alpha}{2}}
		P\left\{ 
		\begin{matrix}
			0&1&\infty&\\
			0&0&1+\alpha+\gamma&\frac{z}{x}\\
			-\alpha &-\alpha&1+\alpha + \frac{x\tilde{c}_1 +2h}{\gamma}&\\
		\end{matrix}
		\right\}
	\end{equation}
	From this we read off that 
	\begin{align}
		& a = 1 + \alpha + \gamma, \\
		& b = 1 + \alpha + \frac{2h- x\tilde{c}_2 }{\gamma}, \\
		& c = 1 + \alpha.
	\end{align}
	As a consequence we can solve for $\gamma$, regularity demands
	\begin{equation}
		-\alpha = c - a - b, \;\;\;\;\; \Rightarrow \;\;\;\; \gamma=-\frac{1}{2} + \frac{1}{2}\nu,
	\end{equation}
	where we define
	\begin{equation}
		\nu^2 \equiv 1+ 4x\tilde{c}_2 - 8h.
		\label{nudefinition}
	\end{equation}
	From this we find a basis of solutions that possesses a diagonal monodromy matrix for a small loop around $z=0$
	\begin{equation}
		\rho^{(0)}_{\pm}(z) = \left(\frac{z}{x}\right)^{\frac{1\pm \alpha}{2}}\left(\frac{z}{x}-1\right)^{\frac{1+\alpha}{2}} \,_2F_1\left(\frac{1}{2}(1+\nu +\alpha \pm \alpha), \frac{1}{2}(1-\nu +\alpha \pm \alpha); 1\pm \alpha, \frac{z}{x}\right).
	\end{equation}
	Finding a basis of solutions that diagonalizes the monodromy around $z=x$ is very simple in this case. We apply the transformation $\frac{z}{x} \rightarrow 1- \frac{z}{x}$, which swaps the first two columns of the P-symbol, since these columns are identical the desired basis of solutions is given by
	\begin{equation}
		\rho^{(x)}_{\pm}(z) = \left(1-\frac{z}{x}\right)^{\frac{1\pm \alpha}{2}}\left(-\frac{z}{x}\right)^{\frac{1+\alpha}{2}} \,_2F_1\left(\frac{1}{2}(1+\nu +\alpha \pm \alpha), \frac{1}{2}(1-\nu +\alpha \pm \alpha); 1\pm \alpha, 1-\frac{z}{x}\right).
	\end{equation}    
	These are both two bases of a linear 2nd-order differential equation, as a consequence there should exist a linear transformation that takes one basis of solutions to the other, i.e.
	\begin{equation}
		\begin{pmatrix}
			\rho^{(0)}_{+}(z)\\
			\rho^{(0)}_{-}(z)\\
		\end{pmatrix}
		= C
		\begin{pmatrix}
			\rho^{(x)}_{+}(z)\\
			\rho^{(x)}_{-}(z)\\
		\end{pmatrix}.
	\end{equation}
	We can construct the matrix $C$ through means of the Kummer relations
	\begin{align}
		& \,_2F_1(a,b;c,z)=\frac{\Gamma(a+b-c)\Gamma(c)}{\Gamma(a)\Gamma(b)}(1-z)^{c-a-b}\,_2F_1(c-a,c-b;c-a-b+1,1-z) \nonumber \\
		& +\frac{\Gamma(c-a-b)\Gamma(c)}{\Gamma(c-a)\Gamma(c-b)}\,_2F_1(a,b;a+b-c+1,1-z).
	\end{align}
	So we can express the solutions $\psi^{(0)}_{\pm}$ respectively as
	\begin{equation}
		\rho^{(0)}_{+} = \frac{(-1)^{\alpha}\Gamma(\alpha)\Gamma(1+\alpha)}{\Gamma(\frac{1}{2}+\frac{1}{2}\nu+\alpha)\Gamma(\frac{1}{2}-\frac{1}{2}\nu+\alpha)} \rho^{(x)}_{-}(z) + \frac{(-1)^\alpha \Gamma(-\alpha)\Gamma(1+\alpha)}{\Gamma(\frac{1}{2}-\frac{1}{2}\nu)\Gamma(\frac{1}{2}+\frac{1}{2}\nu)} \rho^{(x)}_{+}(z)
	\end{equation}
	For the $\rho^{(0)}_{-}$ solution we require the Euler identity
	\begin{equation}
		\,_2F_1(a,b;c,z) = (1-z)^{c-a-b}\,_2F_1(c-a,c-b;c,z),
	\end{equation}
	which results in
	\begin{equation}
		\rho^{(0)}_{-} = -\frac{(-1)^{-\alpha}\Gamma(\alpha)\Gamma(1-\alpha)}{\Gamma(\frac{1}{2}+\frac{1}{2}\nu)\Gamma(\frac{1}{2}-\frac{1}{2}\nu)}\rho^{(x)}_{-}(z) - \frac{(-1)^{\alpha} \Gamma(-\alpha)\Gamma(1-\alpha)}{\Gamma(\frac{1}{2}+\frac{1}{2}\nu-\alpha)\Gamma(\frac{1}{2}-\frac{1}{2}\nu-\alpha)}\rho^{(x)}_{+}(z).
	\end{equation}
	By applying the reflection formula
	\begin{equation}
		\Gamma(z)\Gamma(1-z) = \frac{\pi}{\sin(\pi z)},
	\end{equation}
	we find
	\begin{equation}
		C = 
		\begin{pmatrix}
			-\frac{(-1)^{\alpha}\cos(\frac{1}{2}\pi\nu)}{\sin(\pi \alpha)} & \frac{(-1)^{\alpha}\alpha \Gamma(\alpha)^2}{\Gamma(\frac{1}{2} +\frac{1}{2}\nu +\alpha)\Gamma(\frac{1}{2} -\frac{1}{2}\nu +\alpha)} \\
			\frac{-(-1)^{\alpha}\alpha \Gamma(-\alpha)^2}{\Gamma(\frac{1}{2} +\frac{1}{2}\nu -\alpha)\Gamma(\frac{1}{2} -\frac{1}{2}\nu -\alpha)} & \frac{(-1)^{\alpha}\cos(\frac{1}{2}\pi\nu)}{\sin(\pi \alpha)} 
		\end{pmatrix}.
	\end{equation}
	The monodromy matrix of a cycle surrounding both the singularities at 0 and $x$ is given by
	\begin{equation}
		M_{0,x} = MCMC^{-1},
		\label{leadingmonodromymatrix}
	\end{equation}
	see fig. \ref{monodromycontour}. Here $M$ is given by
	\begin{equation}
		M = 
		\begin{pmatrix}
			-e^{i\pi \alpha} & 0\\
			0 & -e^{-i\pi \alpha}
		\end{pmatrix}.
	\end{equation}
	We can impose the constraint that $M_{0,x}$ conjugate to the matrix \eqref{characteristicmonodromy}, the monodromy around a single regular point with scaling weight $H_p = \frac{c}{6}h_p$, i.e. with
	\begin{equation}
		\alpha_p = \sqrt{1-4 h_p }.
	\end{equation}
	I.e. we will project out the conformal block with intermediate conformal primary $H_p$. The trace of this monodromy cycle is therefore constrained to
	\begin{equation}
		\text{Tr}(M_{0,x}) = -2\cos(\pi \alpha_p).
		\label{traceconstraint}
	\end{equation}
	We can compute the trace
	\begin{equation}
		\text{Tr}(M_{0,x}) = \text{Tr}(MCMC^{-1}) = -2 \cos(\pi \nu)
	\end{equation}
	Note the following consistency check, if we reverse the contour, contracted it around infinity and used the indicial equation \eqref{indicialequation} for the scaling weight at infinity \eqref{weigthatinfinity} we would have obtained the same result. Using \eqref{traceconstraint} to solve for $\nu$ results in
	\begin{equation}
		\nu = \alpha_p.
	\end{equation}
	Reminding ourselves that 
	\begin{equation}
		\nu^2 = 1 + 4x \tilde{c}_2 - 8h,
	\end{equation}
	we can solve for the transformed accessory parameter
	\begin{equation}
		\tilde{c}_2 = \frac{2h - h_p}{x}.
	\end{equation}
	To find the semi-classical conformal block $f(x,H)$ we remind ourselves that $\tilde{c_2}$ is defined as $\tilde{c_2}= \partial f/\partial x$. By integrating this expression with respect to $x$ we find that up to an irrelevant constant term
	\begin{equation}
		f(x,H) = (2h-h_p)\log(x).
	\end{equation}
	From which we find, as dictated by \eqref{conformalblock}, that the leading part of the conformal block is given by
	\begin{equation}
		\mathcal{F}(x,H) = e^{-\frac{c}{6}f(x,H)} = x^{-2H+H_p}.
	\end{equation}
	Hence to leading order we find that the conformal block just reproduces the overall Frobenius factor of the conformal block.

	\section{The first subleading correction}
	\label{sec:firstcorrection}
	We will now attempt to construct the first subleading correction to the solution of the full Fuchsian equation. Decompose the solution to the Fuchs equation as
	\begin{equation}
		\psi(z) = \rho(z) + x \eta(z) + \mathcal{O}(x^2)
	\end{equation}
	where
	\begin{equation}
		\psi''(z) + T(z)\psi(z) = 0
	\end{equation}
	and
	\begin{equation}
		\rho''(z) + T_0(z)\rho(z) = 0.
	\end{equation}
	Where the leading stress tensor $T_0(z)$ and its correction term $V(z) = T_0(z)-T(z)$ are given by equations \eqref{leadingstresstensor} and \eqref{correctionpotential}. Now knowing that to leading order $\tilde{c}_2 = (2h-h_p)/x$, we can use the accessory parameter constraint to find that to leading order 
	\begin{equation}
		\tilde{c}_3 = 2h + h_p,
	\end{equation}
	thus the dominant part of the correction term $V(z)$ is given by 
	\begin{equation}
		V_0(z) \equiv \frac{2h+h_p}{z(z-1)}.
	\end{equation}
	From this we find that the first subleading correction $\eta(z)$ is subject to the equation
	\begin{equation}
		\eta''(z) + T_0(z)\eta(z) = \frac{1}{x}V_0(z)\rho(z).
		\label{subleadingFuchs}
	\end{equation}
	This is an inhomogeneous version of the same Fuchsian equation that determined the leading order solution $\rho(z)$. For this reason we can express the subleading correction $\eta(z)$ in terms of the leading solutions through means of variation of parameters.
	
	Starting with the basis of solutions $\rho_{\pm}^{(0)}$. Promoting the coefficients of the general solution to functions of $z$ we find
	\begin{equation}
		\eta^{(0)}_{\pm} (z) = A_{\pm}^{(0)}(z) \rho_{+}^{(0)} + B_{\pm}^{(0)}(z) \rho_{-}^{(0)}
	\end{equation}
	we find that the functions $\eta^{(0)}_{\pm}$ solve the differential equation
	\begin{equation}
		\eta^{(0)''}_{\pm}(z) + T_0(z)\eta^{(0)}_{\pm}(z) = \frac{1}{x}V_0(z)\rho^{(0)}_{\pm}(z),
		\label{subleadingFuchs2}
	\end{equation}
	Under the conditions that the derivatives of $A_{\pm}^{(0)}(z)$ and $B_{\pm}^{(0)}(z)$ are respectively given by
	\begin{align}
		& A_{\pm}^{(0)'}(z) = -\frac{1}{x} \frac{1}{W} \frac{2h+h_p}{(x-z)(z-1)} \rho_{-}^{(0)} \rho_{\pm}^{(0)},\\  
		& B_{\pm}^{(0)'}(z) = \frac{1}{x} \frac{1}{W} \frac{2h+h_p}{(x-z)(z-1)} \rho_{+}^{(0)} \rho_{\pm}^{(0)},
	\end{align}
	where $W$ is the Wronskian determinant given by
	\begin{equation}
		W = \rho^{(0)}_{+} \rho^{(0)'}_{-} - \rho^{(0)}_{-} \rho^{(0)'}_{+}. 
	\end{equation}

	\subsection{Subleading correction to the monodromy matrix}
	We will now compute the first subleading correction to the trace of the monodromy matrix. As before we will consider a contour $\mathcal{C}$ that encloses the singular points at $z=0$ and $z=x$. We remind ourselves that the corrected basis of solutions is given by
	\begin{equation}
		\psi_{\pm}(z) = \rho_{\pm}(z) + x\eta_{\pm}(z).
	\end{equation}
	\textit{Note:} for the upcoming discussion we will exclusively use the $\rho^{(0)}_{\pm}$ basis for the leading solutions so will suppress the superscript to avoid notational clutter. 
	
	The subleading corrections can be explicitly expressed as
	\begin{equation}
		\eta_{\pm}(z) =  A_{\pm}(z) \rho_{+} + B_{\pm}(z) \rho_{-}.
	\end{equation}
	If we define the matrix elements of the leading order monodromy matrix as
	\begin{equation}
		M_{0,x} = 
		\begin{pmatrix}
			K_{++}(\nu, \alpha) & K_{+-}(\nu, \alpha) \\
			K_{-+}(\nu, \alpha) & K_{--}(\nu, \alpha)
		\end{pmatrix},
	\end{equation}
	then we can explicitly write the transformation rule for the corrected solutions as respectively
	\begin{equation}
		\psi_{+} \rightarrow (1+ x(A_{+}(z) + M_{A_+}))(K_{++} \rho_+ + K_{+-}\rho_-) + x(B_+(z)+M_{B_+})(K_{-+}\rho_+ + K_{--}\rho_-),
		\label{subleadingtrafo1}
	\end{equation}
	and
	\begin{equation}
		\psi_{-} \rightarrow (1+ x(B_{-}(z) + M_{B_-}))(K_{-+} \rho_+ + K_{--}\rho_-) + x(A_-(z)+M_{A_-})(K_{++}\rho_+ + K_{+-}\rho_-).
		\label{subleadingtrafo2}
	\end{equation}
	Where $M_{A_{\pm}}$ and $M_{B_{\pm}}$ are the branch discontinuities of the of the coefficients which are given by the contour integrals
	\begin{equation}
		M_{A_{\pm}} = \oint_{\mathcal{C}} A_{\pm}'(z')dz'.
	\end{equation}
	and
	\begin{equation}
		M_{B_{\pm}} = \oint_{\mathcal{C}} B_{\pm}'(z')dz',
	\end{equation}
	where the integration contour $\mathcal{C}$ is a cycle that encloses the points around $z=0$ and $z=x$ and is assumed to stay within at most an order unity multiple of $x$ radius around the origin. As an explicit example $\mathcal{C}$ could the circle around the origin of radius $2x$. From expressions \eqref{subleadingtrafo1} and \eqref{subleadingtrafo2} we can read off the corrected diagonal elements of the monodromy matrix
	\begin{align}
		& \begin{pmatrix}
			\rho^{+}(z) + x \eta^{+}(z)\\
			\rho^{+}(z) + x \eta^{-}(z)
		\end{pmatrix}
		\rightarrow 
		\tilde{M}_{0,x}\begin{pmatrix}
			\rho^{+}(z) + x \eta^{+}(z)\\
			\rho^{+}(z) + x \eta^{-}(z)
		\end{pmatrix} = \nonumber \\ 
		&\begin{pmatrix}
			K_{++} + x(M_{A_+} K_{++} + M_{B_+} K_{-+}) & ...\\
			... & K_{--} + x(M_{B_-} K_{--} + M_{A_-} K_{+-})
		\end{pmatrix}
		\begin{pmatrix}
			\rho^{+}(z) + x \eta^{+}(z)\\
			\rho^{+}(z) + x \eta^{-}(z)
		\end{pmatrix}.
	\end{align}
	As a consequence the trace of the monodromy matrix including the first subleading correction is given by
	\begin{equation}
		\text{Tr}(\tilde{M}_{0,x}) = -2\cos(\pi\nu) + x (2h+h_p) g(h,h_p),
	\end{equation}
	where we have defined
	\begin{equation}
		g(h,h_p) = \frac{1}{2h+h_p}\left(M_{A_+}K_{++} + M_{B_+}K_{-+} + M_{B_-}K_{--} + M_{A_-}K_{+-}\right).
		\label{tracemodification}
	\end{equation}
	The coefficients $K_{\pm\pm}$ are given in closed form as the matrix elements of \eqref{leadingmonodromymatrix}. The integrands that define $M_{A_{\pm}}$ and $M_{B_{\pm}}$ are quite complex, providing closed form expressions for the integrals is beyond the scope of our ability. Since the $x$-dependence has been filtered out numerical integration is an option.

	\subsection{Variation of the accessory parameter}
	As was seen in the directly preceding section, if we include the subleading correction to the solutions of the Fuchsian ODE than we also find a subleading correction to the trace of the monodromy matrix. In this section we will introduce a first-order variation to the leading solution of the accessory parameter. We will constrain this variation by demanding that it will off-set the newly-introduced correction to the monodromy trace.
	
	As discussed in section \ref{sec:fourheavyoperators}, to leading order the trace of the monodromy constraint gave us an algebraic equation that fixes the accessory parameter
	\begin{equation}
		-2\cos(\pi \alpha_p) = -2\cos(\pi \nu),
		\label{traceconditionexchangeprimary}
	\end{equation}
	where $\alpha_p$ is given by the exchange primary scaling dimension $h_p$ through
	\begin{equation}
		\alpha_p = \sqrt{1-4h_p} \equiv \sqrt{1-24\frac{H_p}{c}}.
	\end{equation}
	In this case we find that 
	\begin{equation}
		\tilde{c}_2^{(0)} = \frac{2}{x}(h-\frac{1}{2}h_p).
		\label{leadingc1}
	\end{equation}
	Including the first subleading correction to the solutions of the Fuchsian ODE then \eqref{traceconditionexchangeprimary} gets modified to
	\begin{equation}
		-2\cos(\pi \alpha_p) = -2\cos(\pi \nu) + x (2h+h_p) g(h, h_p),
		\label{traceconditionexchangeprimarycorrected}
	\end{equation}
	where $g(h,hp)$ is given by \eqref{tracemodification}. We will introduce a small variation to the accessory parameter
	\begin{equation}
		\tilde{c}_2 = \tilde{c}_2^{(0)} + \tilde{c}_2^{(1)}.
	\end{equation}
	We can now attempt to solve for the variation of the accessory parameter $\tilde{c}_2^{(1)}$. We remind ourselves that $\nu$ is given by
	\begin{equation}
		\nu^2 = 1 + 4x \tilde{c}_2 - 8h.
	\end{equation}
	In which we find that the solution for $\tilde{c}_2^{(1)}$ is given by
	\begin{equation}
		\tilde{c}_2^{(1)} = -\frac{\alpha_p (2h+h_p)g(h,h_p)}{4\pi \sin(\pi \alpha_p)}.
		\label{firstvariationc1}
	\end{equation}
	Since $g(h, h_p)$ is independent of $x$ we can integrate with respect with $x$. Combining \eqref{leadingc1} and \eqref{firstvariationc1} and integrating with respect to $x$ we find that 
	\begin{equation}
		f(x,H,H_p) = \frac{6}{c}(2H-H_p)\log(x) + \frac{3}{2c}\frac{\alpha_p (2H+H_p)g(h,h_p)}{\pi \sin(\pi \alpha_p)}x.
	\end{equation}
	By taking the exponent we can now find the first-order corrected version of the conformal block with exchange primary $H_p$
	\begin{equation}
		\mathcal{F}_p(x,H, H_p) = x^{-2H+H_p}\times e^{ \frac{\alpha_p (2H+H_p)g(h,h_p)}{4\pi \sin(\pi \alpha_p)}x}.
		\label{firstorderdressedblock}
	\end{equation}
	Here we have explicitly factored out the dressing factor due to the first subleading correction. Note that the combination $\frac{2H+H_p}{4\pi \sin{\pi\alpha_p}}\alpha_p$ is manifestly real positive as long as $H_p<c/24$. As a consequence the qualitative behavior of the dressing factor of the conformal block as a function of $x$ is controlled by the signs and relative magnitudes of the real and imaginary parts of $g(h,h_p)$.

	\subsection{Special case: the identity block}
	The identity block is a special case, if $h_p=0$ then $\alpha_p=1$ and subsequently $\sin(\pi\alpha_p)$ vanishes identically. If we follow the same steps as in the preceding section we find instead the following expression for the first-order variation of the accessory parameter
	\begin{equation}
		\tilde{c}_2^{(1)} = \frac{1}{2\pi} \sqrt{\frac{2h}{x} g(h,0)}.
	\end{equation}
	This point introduces another subtlety, $g(h,h_p)$ is given by
	\begin{equation}
		g(h,h_p) = \frac{1}{2h}\left(M_{A_+}K_{++} + M_{B_+}K_{-+} + M_{B_-}K_{--} + M_{A_-}K_{+-}\right),
	\end{equation}
	when $h_p=0$ then $K_{+-} = K_{-+} = 0$ and the contour integrals $M_{A_+}$ and $M_{B_-}$ vanish identically. As such we find that to our order of approximation the first subleading correction vanishes identically and one would have to go to the next order in the solutions to the Fuchsian ODE.

	\subsection{Numerical integration of $g(h,h_p)$}
	The goal of our computation was to compute the overlap of the initial state with itself at Lorentzian time $t=0$. In the introduction section we had established that we can relatively straightforwardly compute the Lorentzian time-evolution by analytically continuing the real parameter $\sigma$ to the complex upper half-plane. 
	
	In our conformal frame it is not $\sigma$ that enters the expressions but the cross ratio $x = \sigma^2$, the effect of continuing $\sigma$ is easily explored
	\begin{equation}
		x = \sigma^2 \Longrightarrow (\sigma + it)^2 = x + 2i\sigma t - t^2.
	\end{equation}
	Hence we see that after an initial time-scale of characteristic length $\text{Re}(\sigma)$ that the Lorentzian time dependence is determined by the term $-t^2$. From the dressing factor contribution that comes from the first-order correction to the conformal block \eqref{firstorderdressedblock} we see that the qualitative behavior of our transition amplitude as we advance $t$ is given by the phase of $g(h,h_p)$ which is given in \eqref{tracemodification}. While the matrix elements $K_{\pm\pm}$ are known exactly, the factors $M_{A_{\pm}}$ and $M_{B_{\pm}}$ are given by a contour integral that surround two branch cuts of a complex integrand given by a ratio of a linear combination of products of hypergeometric functions. Analytical integration is both outside of our scope and capability, but we can make some progress numerically. In principle the integration contour can be deformed as long as we do not cross any additional regular singular points, in practice our integrand is only valid when $|z|$ is not significantly larger than $x$. In order to avoid ambiguities we will be explicit and choose our integration contour $\mathcal{C}$ to be the circle with radius $2x$.
	
\begin{figure}
		\centering
		\includegraphics[width=1\linewidth]{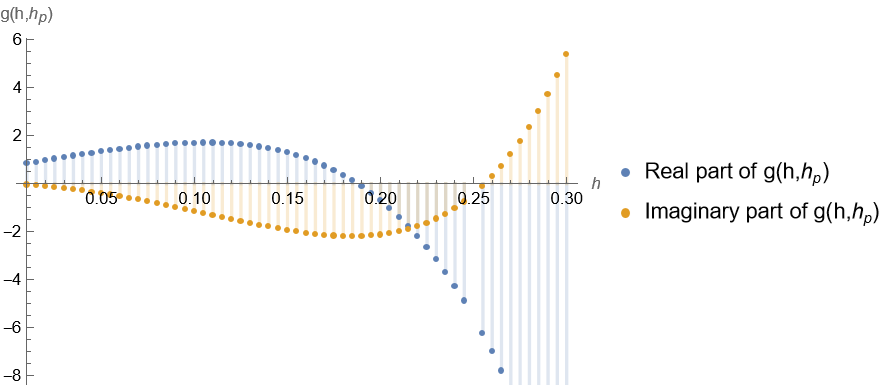}
		\caption{A numerical plot of $g(h,h_p)$ as a function of $h$ with fixed $h_p = \frac{1}{50}$. Note the sign flip of the real part, which occurs earlier than at the vanishing of $\alpha$ at $h=1/4$.}
		\label{numericalintegrationhp1over50}
	\end{figure}

	For fixed $h_p = 1/50$ the result for the real and imaginary parts of $g(h,h_p)$ as a function of $h$ are displayed in fig. \ref{numericalintegrationhp1over50}
	
\begin{figure}
		\centering
		\includegraphics[width=1\linewidth]{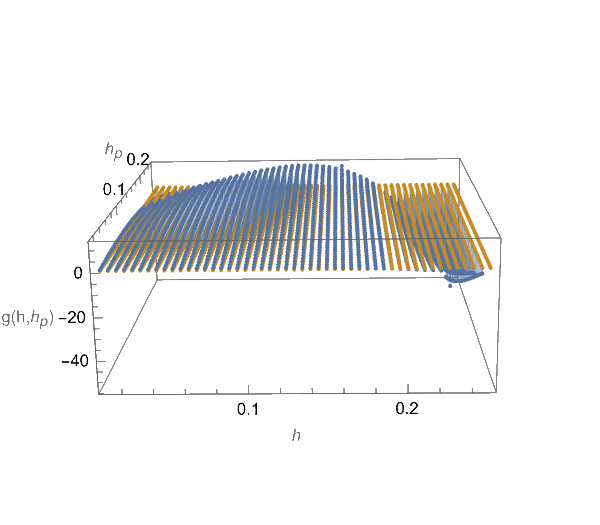}
		\caption{A numerical 3d plot of the real part $g(h,h_p)$ as a function of $h$ and $h_p$. The orange plane is the zero plane. Note the sign flip of the real part appears to be independent of $h_p$, suggestive that the transition is a property of the CFT state and not the intermediate primary.}
		\label{numericalintegration3d}
	\end{figure}

	We can see that there is a sign flip that occurs in the real part of $g(h,h_p)$ this suggests that there is a value for $h$ where there exists a transition in the scattering amplitude as a function of time. There exists a point where the scattering amplitude transitions from a growing oscillating function of $t$ to a decaying function of $t$. In addition, in the 3d plot of the real part of $g(h,h_p)$ as a function of both $h$ and $h_p$ we find that the point of the transition is independent of $h_p$, see fig. \ref{numericalintegration3d}.

	This suggests that the transition is an inherent property of the CFT state as it is independent of the choice of particular conformal block, given a fixed OPE channel.

	\section{Discussion}
	We have selected a range of parameters in order to, to best approximation, mimic the environment of astrophysical-scale black hole physics in bulk AdS. By choosing the central charge to be large we have selected the AdS radius to be parametrically larger than the Plank length. By choosing heavy scaling dimension $H\propto c$ we have elected to release initial objects at the boundary with masses off the order of magnitude of black holes (but below the BTZ mass gap). By taking into account only the leading order contribution in $c$ in the CFT we have effectively turned of quantum corrections in the bulk. By taking $\sigma$ to be small we have created local excitations at the AdS boundary, but by making sure $\sigma$ is not too small we have restricted the initial energy of our system to sub-Planckian scales.
	
	In the process we found that every conformal block in our channel, i.e. all values of $h_p$, predict that there is a critical value for $h$ where there is a transition point from a growing oscillating function of $t$ to a decaying one. We interpret this transition as a shift from a phase where the intermediate channel is dominated by elastic scattering to a phase dominated by an intermediate black hole. 
	
	We interpret this as a reflection of the general expectation that if two colliding large objects fail to undergo black hole collapse they will continue to oscillate around the center of AdS forever. Whereas if they do collapse, the intermediate decay channel that would lead to two-into-two scattering would involve a black hole decay process where the black hole spontaneously radiates its original two constituents from which it was formed. By time-reversal symmetry this is strictly speaking an allowed process but on that is very heavily suppressed. This is in accordance with the witnessed exponential decay of its transition amplitude.

	\section*{Acknowledgements}
	The authors have benefited greatly from various discussions with various people, in particular Kyriakos Papadodimas, Tom\'as Proch\'azka, Shiraz Minwalla, Joris Raeymaekers and Igor Khavkine. The work of GV was in part supported by the Grant Agency of the Czech Republic under the grant EXPRO 20-25775X and in part supported by a KIAS Individual Grant (PG100401) at Korea Institute for Advanced Study.

	\appendix
	
	\section{The leading-order solutions in terms of associated Legendre functions}
	In the main body of the text it was useful to maintain the form of the leading solutions $\rho^{(0)}_{\pm}(z)$ and $\rho^{(x)}_{\pm}(z)$ in terms of hypergeometric functions as it made it explicitly clear how to relate the two bases through means of the Kummer relations. Outside of that, we can utilize the implicit symmetry induced by the identical weights of the regular singular points to simplify the solutions. By reducing the solutions to associated Legendre functions we can make the notation more compact at the cost of obscuring some of the branch structure. We manage this by applying the identity
	\begin{equation}
		\,_2F_1(\alpha, 1-\alpha; c, z) = \Gamma(c) z^{\frac{1-c}{2}} (1-z)^{\frac{c-1}{2}} P_{-\alpha}^{1-c}\left(1-2\frac{z}{x}\right),
	\end{equation}
	where $P_{a}^{b}(z)$ is the associated Legendre P-function, to simplify the $-$-solutions to
	\begin{equation}
		\rho_{-}^{(0)}(z) = (-1)^{-\frac{\alpha}{2}}\,\sqrt{\frac{z}{x}}\sqrt{\left(\frac{z}{x}-1\right)}\; \Gamma(1-\alpha) P^{\alpha}_{-\frac{1}{2}-\frac{1}{2}\nu}\left(1-2\frac{z}{x}\right).
	\end{equation}
	For the $+$-solution we will apply the Euler identity mentioned above
	\begin{equation}
		\,_2F_1(a,b;c,z) = (1-z)^{c-a-b}\,_2F_1(c-a,c-b;c,z),
	\end{equation}
	which results in
	\begin{equation}
		\rho^{(0)}_{+}(z) = (-1)^{\frac{\alpha}{2}}\sqrt{\frac{z}{x}}\sqrt{\frac{z}{x}-1}\; \Gamma(1+\alpha) P^{-\alpha}_{-\frac{1}{2}+\frac{1}{2}\nu}\left(1-2\frac{z}{x}\right).
	\end{equation}
	Exploiting the linearity of the original ODE to strip off extraneous factors results in the following basis of solutions
	\begin{equation}
		\rho^{(0)}_{\pm}(z) = \sqrt{z}\sqrt{z-x} \; P^{\mp \alpha}_{-\frac{1}{2}\pm \frac{1}{2}\nu}\left(1-2\frac{z}{x}\right).
	\end{equation}
	By repeating the exact same method we can also reduce the $\psi^{(x)}_{\pm}(z)$ basis to the basis
	\begin{equation}
		\rho^{(x)}_{\pm}(z) = \sqrt{z}\sqrt{z-x} P^{\mp \alpha}_{-\frac{1}{2} \pm \frac{1}{2}\nu}\left(2\frac{z}{x}-1\right) 
	\end{equation}
	
	%At this level, the only potential distinct change we witness is a sign change at 
	%\begin{equation}
	%8 \frac{6H}{c} = 3 \;\;\;\;\; \Rightarrow \;\;\;\;\; \frac{H}{c} = \frac{1}{16}.
	%\end{equation}
	%Which is close but not the same as \cite{Kusuki:2018nms} which finds a transition at $H/c = 1/32$.

	\bibliography{references}
	\bibliographystyle{JHEP}

\end{document}